\documentstyle[aas2pp4]{article}

\begin{document}

\title{
The Influence of Parameters on the Flow Structure in Semidetached Binary
Systems: 3D Numerical Simulation}

\author{D.V.Bisikalo\altaffilmark{1}}
\affil{Institute of Astronomy of the Russian Acad. of Sci.,
Moscow, Russia}

\author{A.A.Boyarchuk}
\affil{Institute of Astronomy of the Russian Acad. of Sci.,
Moscow, Russia}

\author{O.A.Kuznetsov\altaffilmark{2}}
\affil{Keldysh Institute of Applied Mathematics, Moscow, Russia}

\author{V.M.Chechetkin}
\affil{Keldysh Institute of Applied Mathematics, Moscow, Russia}

\altaffiltext{1}{\large E-mail address: {\it bisikalo@inasan.rssi.ru}}
\altaffiltext{2}{\large E-mail address: {\it kuznecov@spp.keldysh.ru}}

\begin{abstract}
    The basic parameters determining the flow pattern for a
nonviscous, non-heat-conducting gas in a semidetached binary
system without a magnetic field are identified.
Three-dimensional gas--dynamical modeling of the mass transfer
enables investigation of the influence of these parameters on
the structure of gas flows. The parameter on which the flow
pattern depends most strongly is the adiabatic index $\gamma$.
The effect of other parameters is small, and only leads to
unimportant quantitative changes in the solutions obtained. The
main properties of flows typical of semidetached binaries
without magnetic fields are summarized.
\end{abstract}

\section{Introduction}

    Semidetached binary systems belong to the class of
interacting stars, in which one of the components fills its
Roche lobe and there is mass exchange between the components
through the inner Lagrange point. Beginning with the pioneering
work of Prendergast [1], the gas dynamics of matter flows in
semidetached systems have been extensively investigated by many
authors.  As a rule, two-dimensional (2D) numerical models have
been used in calculations (see, for example, [2--9]).
Unfortunately, the limitations of the 2D approach cast doubt on
the adequacy of some results obtained using these models.
Analytical studies of the applicability of 2D models for
accretion disks [10] have shown that these models give correct
results only in certain limited cases (with an adiabatic index
$\gamma=1$ for disks in an external gravitational field and
$\gamma=2$ for self-gravitating disks). In all other cases, a
three-dimensional (3D) treatment must be employed. The
comparison [11] of numerical calculations performed using 2D
and 3D models confirmed the limited applicability of 2D
models, and demonstrated the necessity of using 3D models in
numerical analyses of the flow structures in semidetached
binaries.

    Three-dimensional modeling calls for substantial
computational resources, and for this reason, such studies are
few in number (the best known are [12--21]).  In addition, the
morphology of the flows in semidetached binaries is examined in
only a few of these, since many of these studies were performed
in a restricted formulation in which the process of
establishment of flow formation was not taken into
consideration.  The variety of binary systems studied and
approaches used makes it impossible to generalize the results
obtained, or to make at least qualitative predictions about the
flow structures expected for semidetached systems. At the same
time, such studies are extremely important, since data on the
structure of gas flows are necessary for interpretation of the
observational data, and the time-consuming nature of 3D modeling
makes it impossible to compute the flow pattern for each
particular system.

    Here, we consider the results of 3D numerical modeling of
the flow of an ideal gas in a semidetached binary system
assuming the absence of magnetic field and radiative heating and
cooling of the gas. To make the study more systematic, we
generalize the results of eight of our own calculations, and
have also used, in some cases, the results of other studies. We
identified the basic parameters of this problem (Section 2) and
successively examined the effect of each parameter on the flow
structure (Section 3). Features of the flow brought about by the
influence of various parameters and the typical characteristics
of the flow structure are described in Section 4.

\section{Parameters of the Problem}

    In order to perform gas--dynamical studies of the flow
structures in binary systems, it is necessary to specify
values of the parameters describing the system under
consideration and of the parameters used in the mathematical
model used.

\subsection{Parameters of binary system}

    Numerical analyses of the matter flows in binary systems to
not require the specification of all seven orbital elements for
the binary star, since the orbital plane is assumed to be known
in advance. In this case, we can describe the motion of the
system's components using the mass of the donor star $M_1$ and
the mass of the accretor star $M_2$ (or equivalently, the total
mass of the system $M$ and the mass ratio of the components
$q=M_2/M_1$), the semimajor axis of the orbit $A$, the rotation
period $P_{orb}$ (or, equivalently, the angular velocity
$\Omega= 2\pi/P_{orb}$), and the orbital eccentricity $e$.  Note
that the application of Kepler's third law decreases the
required number of parameters, making it possible to use,
instead of the four parameters $M$, $q$, $A$, and $\Omega$, any
three of these to describe the system in a point-mass
approximation. The geometric sizes of the components should also
be included among the parameters describing the system. For the
semidetached systems considered here, we assume that the donor
star fills its Roche lobe and, therefore, its geometric
characteristics are completely specified by the parameters of
the system. The size of the accretor star $R_2$ is an
independent parameter that must be given separately.

\subsection{Model parameters}

    In our numerical model, we use the Euler equations in the
form of integral conservation laws to describe the flow of a
nonviscous, non-heat-conducting gas in a binary system without
magnetic or radiation fields. This permits us to obtain
discontinuous (nonisentropic) solutions satisfying the
Rankine--Hugoniot conditions at the discontinuities (shock waves
and contact discontinuities). We complete the system of
gas--dynamical equations with the equation of state of an ideal
gas with adiabatic index $\gamma$:
$P=(\gamma-1)\rho\varepsilon$, where $P$ is the pressure, $\rho$
is the density, and $\varepsilon$ is the specific internal
energy.

    To solve the system of gas--dynamical equations, we must
specify the boundary conditions at the surfaces of the system's
components and at the surface of the computation region. The
states of the gas in the near-surface layers of the donor and
accretor stars are determined by the gas densities $\rho_1$ and
$\rho_2$, the temperatures $T_1$ and $T_2$ (or, equivalently,
the sound speeds $c_1$ and $c_2$), and the gas-velocity vectors
${\bf v}_1$ and ${\bf v}_2$. The boundary conditions at the
stellar surfaces are obtained by the standard procedure of
solving Riemann problem between the states immediately under the
stellar surface and the first computation node (see, for
example, [2, 3]). A free-outflow condition is specified at the
outer boundary of the computation region.

    Thus, in our formulation, the motion of a nonviscous gas in
a binary system without magnetic and radiation fields depends on
five parameters specifying the binary system:  $q$, $A$,
$\Omega$, $e$, $R_2$ and seven model parameters:  $\gamma$,
$\rho_1$, $c_1$, ${\bf v}_1$, $\rho_2$, $c_2$, ${\bf v}_2$.

    Understanding the necessity of studying successively the
roles of individual parameters, we introduce a number of
additional constraints that decrease the total number of
parameters. In particular, we assume that we can neglect the
stellar wind produced by the accretor star. This is standard in
studies of flows in semidetached binaries, and is widely used in
computations (see, for example, [12--19]). In this case, the
only influence of the accretor on the flow of matter is
gravitational, and we can adopt a condition of complete
accretion at its surface, i.e., the condition that all gas
reaching the surface of this component is absorbed. In order to
be able to perform a stage-by-stage investigation of the effect
of various parameters on the flow structure, we also add a
number of constraints on the type of system under consideration:
we assume that (i) the orbits of the components are circular,
which allows us to fix the parameter $e=0$; (ii) the accretor
is a compact object (a white dwarf, neutron star, or black
hole), so that, when $R_2\ll A$, the general structure of the
flow is determined only by the gravitational field of the
accretor and, accordingly, the parameter $R_2$ can be excluded;
(iii) the rotation of the donor star is synchronized with the
orbital motion of the system, so that, in a reference frame
rotating with the angular velocity of the system, the gas
velocity at the surface of the donor star is directed along the
normal to this surface, and we can replace the vector parameter
${\bf v}_1$ by the scalar parameter $v_1$ (equal to the velocity
component perpendicular to the donor-star surface).

    In the above formulation, the flow structure in the binary
system depends on seven parameters:  $q$, $A$, $\Omega$,
$\gamma$, $\rho_1$, $c_1$, $v_1$.  It is noteworthy that the
solution is not affected by the boundary density at the donor
star, since the system of equations can be scaled with respect
to $\rho$ (and simultaneously with respect to the pressure $P$).
In the numerical model, an arbitrary $rho_1$ value can be used.
When considering a specific system with known mass-loss rate,
however, we must scale the calculated density values by the
ratio of the true and model densities at the surface of the
donor star if we wish to find the real density values for the
system. For convenience in the numerical solution, the original
system of equations can be written in dimensionless form.
Following the usual practice (see, for example, [2, 3]), we use
the distance $A$ between the components and the reciprocal of
the angular rotation velocity of the system $\Omega^{-1}$, to
make the equations dimensionless with respect to space and time,
respectively. In this case, the remaining six parameters
determining the flow structure can be reduced to a combination
of four dimensionless parameters $q=M_2/M_1$,
$\epsilon=c_1/(\Omega A)$, ${\cal M}=v_1/c_1$, $\gamma$, and
solutions obtained for two systems with the same parameter
values will differ only by a scaling factor.

\section{Results}

    To determine the influence of various parameters on the flow
structure, we carried out 3D numerical modeling of several
binary systems. We successively varied each of the parameters
while fixing the values of the remaining parameters. It is
evident that in a full analysis, the number of values considered
for each parameter (in the total characteristic range of values)
must be as large as possible. Unfortunately, 3D modeling
requires substantial computational resources, and we restricted
ourselves to only eight calculations.  The parameters used in
these calculations are summarized in Table~I. The limited
number of calculations makes it impossible to study the
dependence of the results on a selected parameter in detail.
Nevertheless, analysis of our results enables identification of
the parameters most strongly affecting the flow structure, and
determination at a qualitative level of trends in the solution
behavior.

\begin{figure*}[t]
\epsscale{1.6}
\plotone{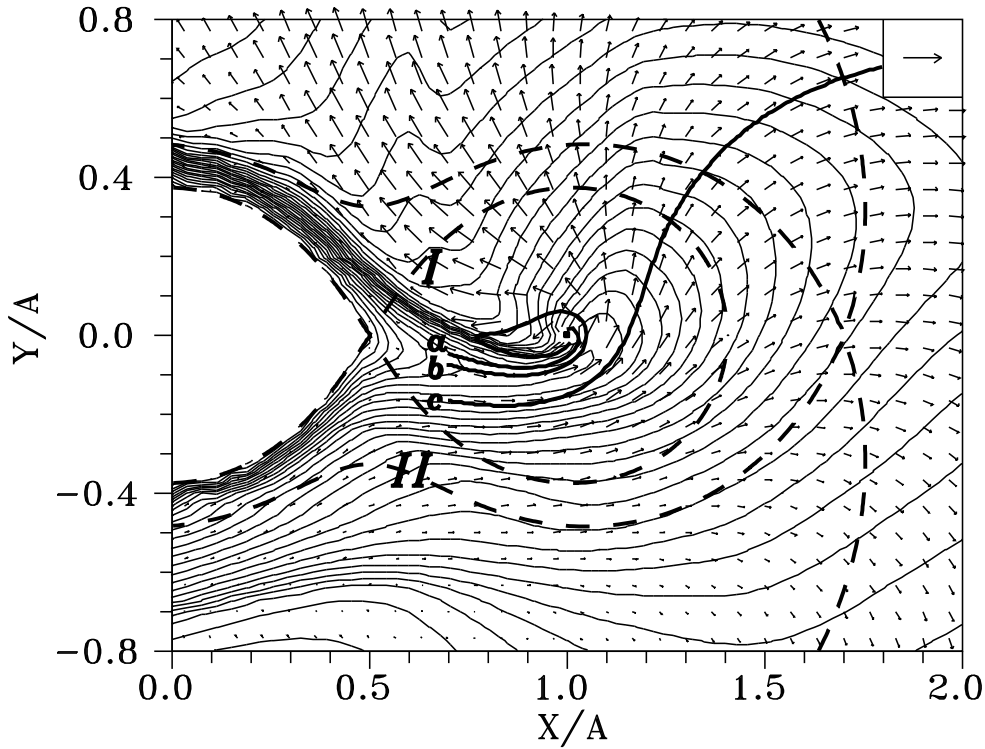}
\caption{(a) density contours and velocity vectors in the
equatorial plane of the binary system for calculation 1
specified in Table~I. The vector in the upper right corner
corresponds to the velocity $4A\cdot\Omega$. Three flowlines,
denoted `$a$', `$b$', and `$c$' illustrate the directions of the
flow of matter in the system. The dashed lines are the Roche
equipotential surfaces.}
\end{figure*}

\renewcommand{\thefigure}{1 -- {\it continued}}
\begin{figure*}[t]
\plotone{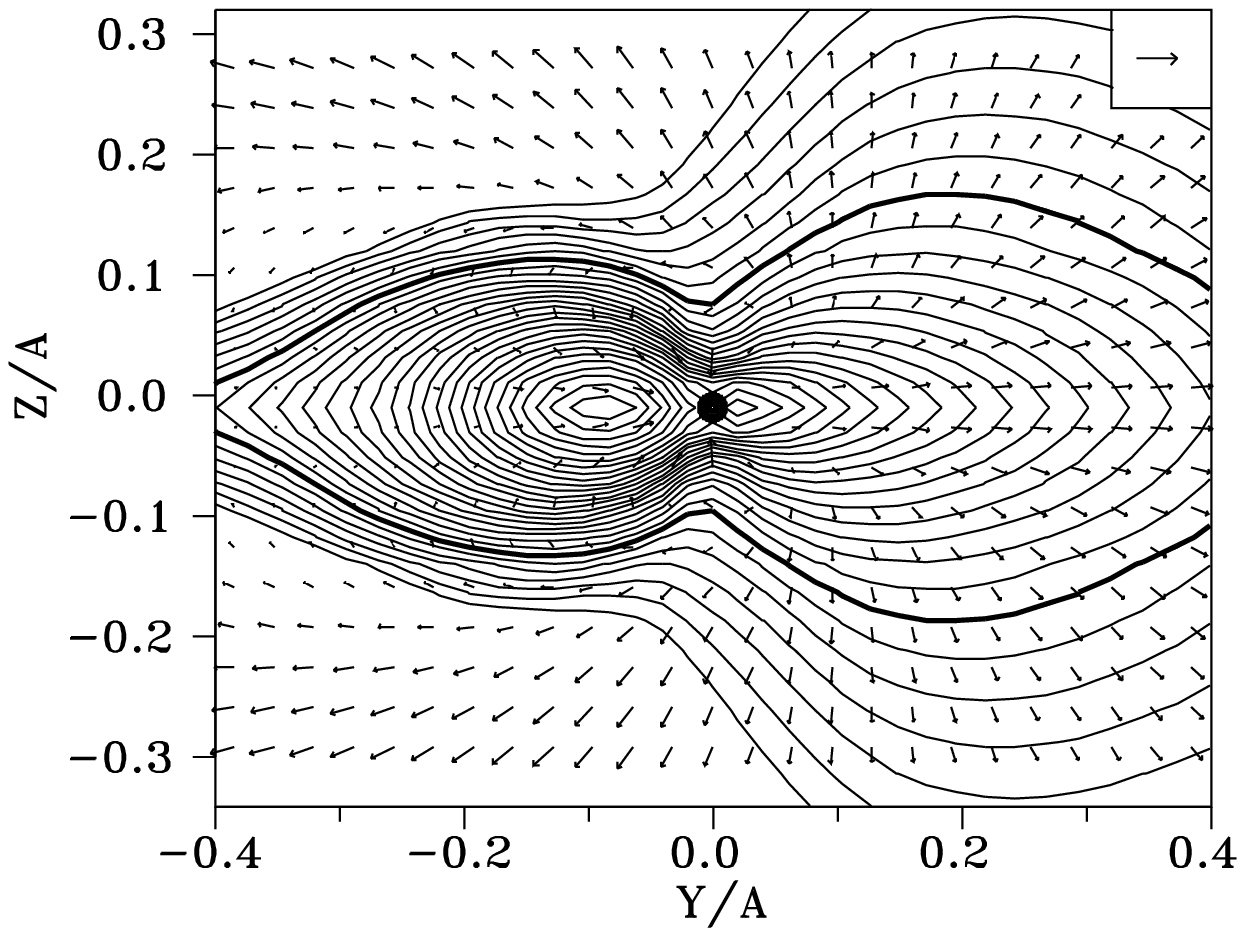}
\caption{(b) density contours and
velocity vectors in the plane passing through the accretor star
and perpendicular to the line connecting the centers of the
system's components. The vector in the upper right corner
corresponds to the velocity $2A\cdot\Omega$. The range of the
density variations is 0.1 -- 0.0003${\rho}_0$. The thick solid line
corresponds to the density 0.001${\rho}_0$.}
\end{figure*}

\renewcommand{\thefigure}{2}
\begin{figure*}[p]
\plotone{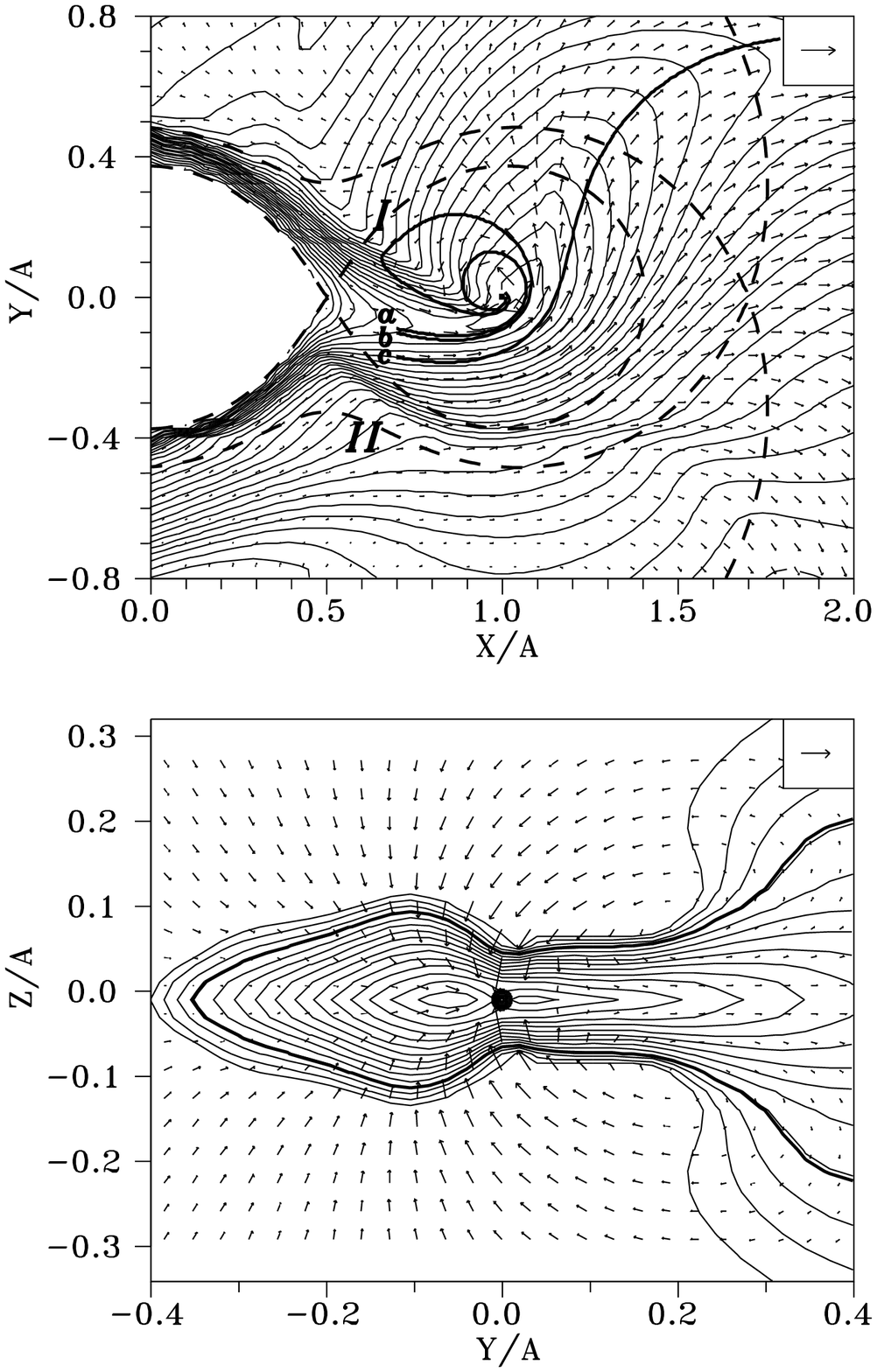}
\caption{Same as Figure~1 for calculation 2 from Table~I.}
\end{figure*}

\renewcommand{\thefigure}{3}
\begin{figure*}[p]
\plotone{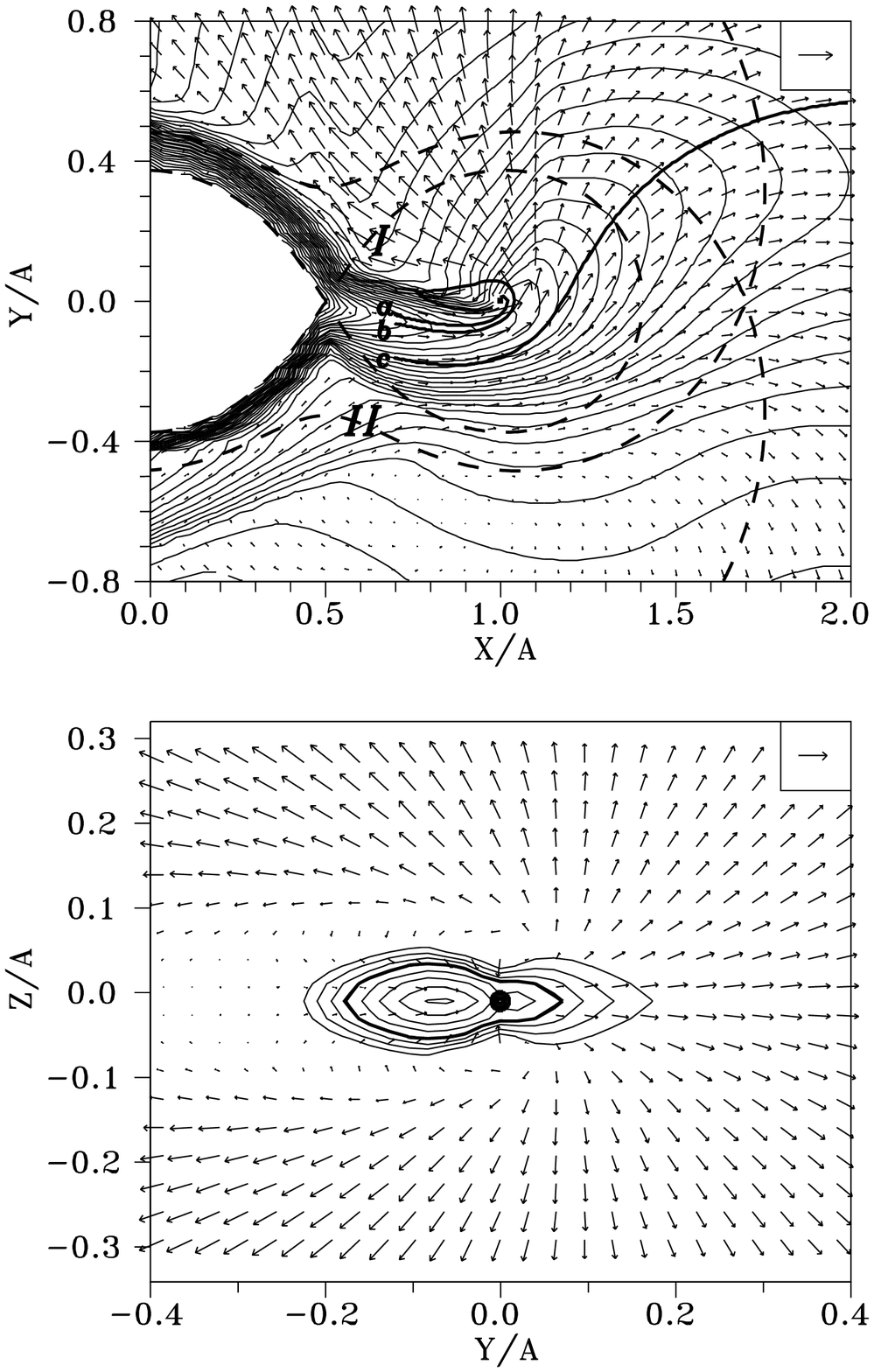}
\caption{Same as Figure~1 for calculation 3 from Table~I.}
\end{figure*}

\renewcommand{\thefigure}{4}
\begin{figure*}[p]
\plotone{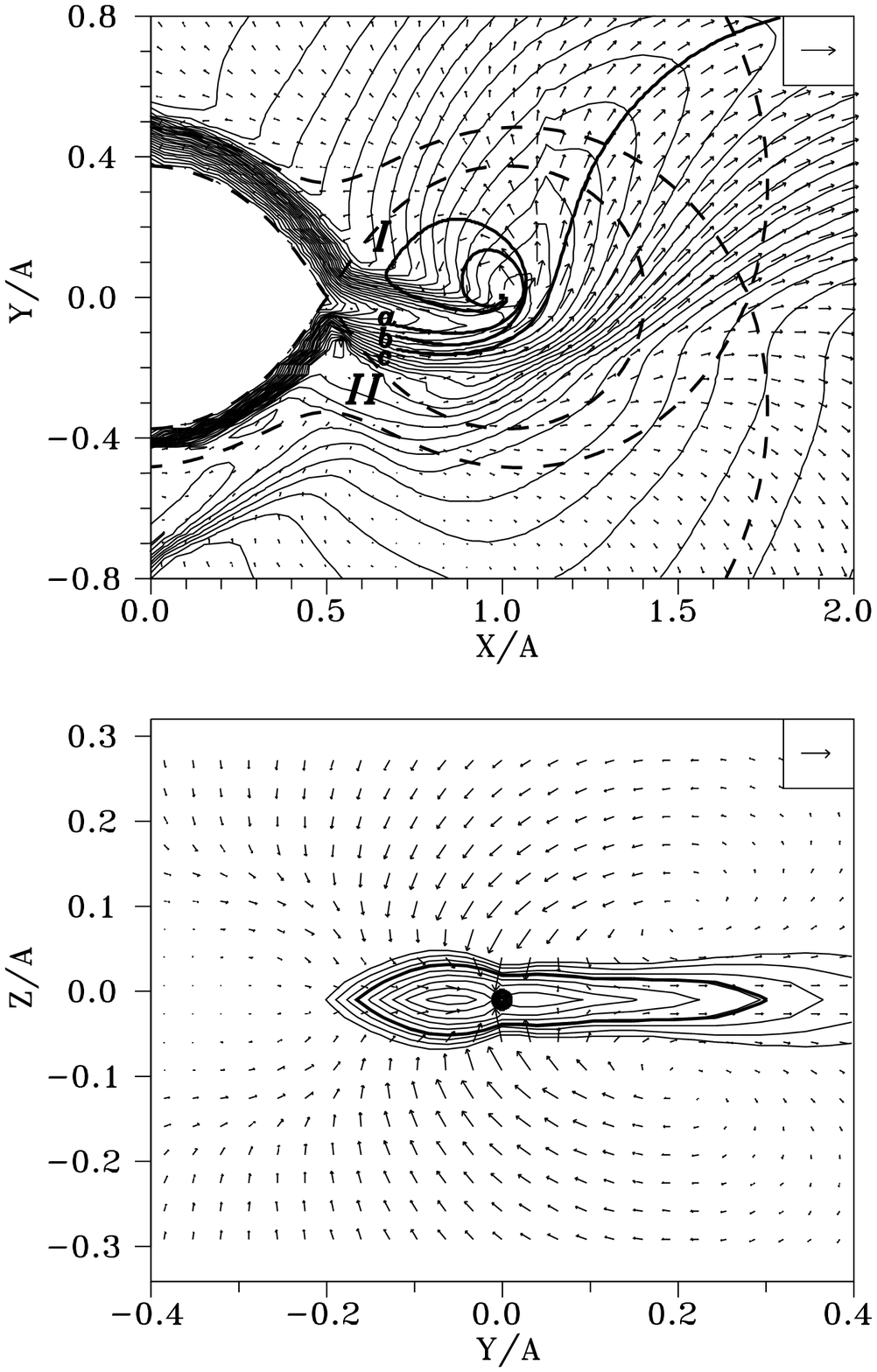}
\caption{Same as Figure~1 for calculation 4 from Table~I.}
\end{figure*}

\renewcommand{\thefigure}{5}
\begin{figure*}[p]
\plotone{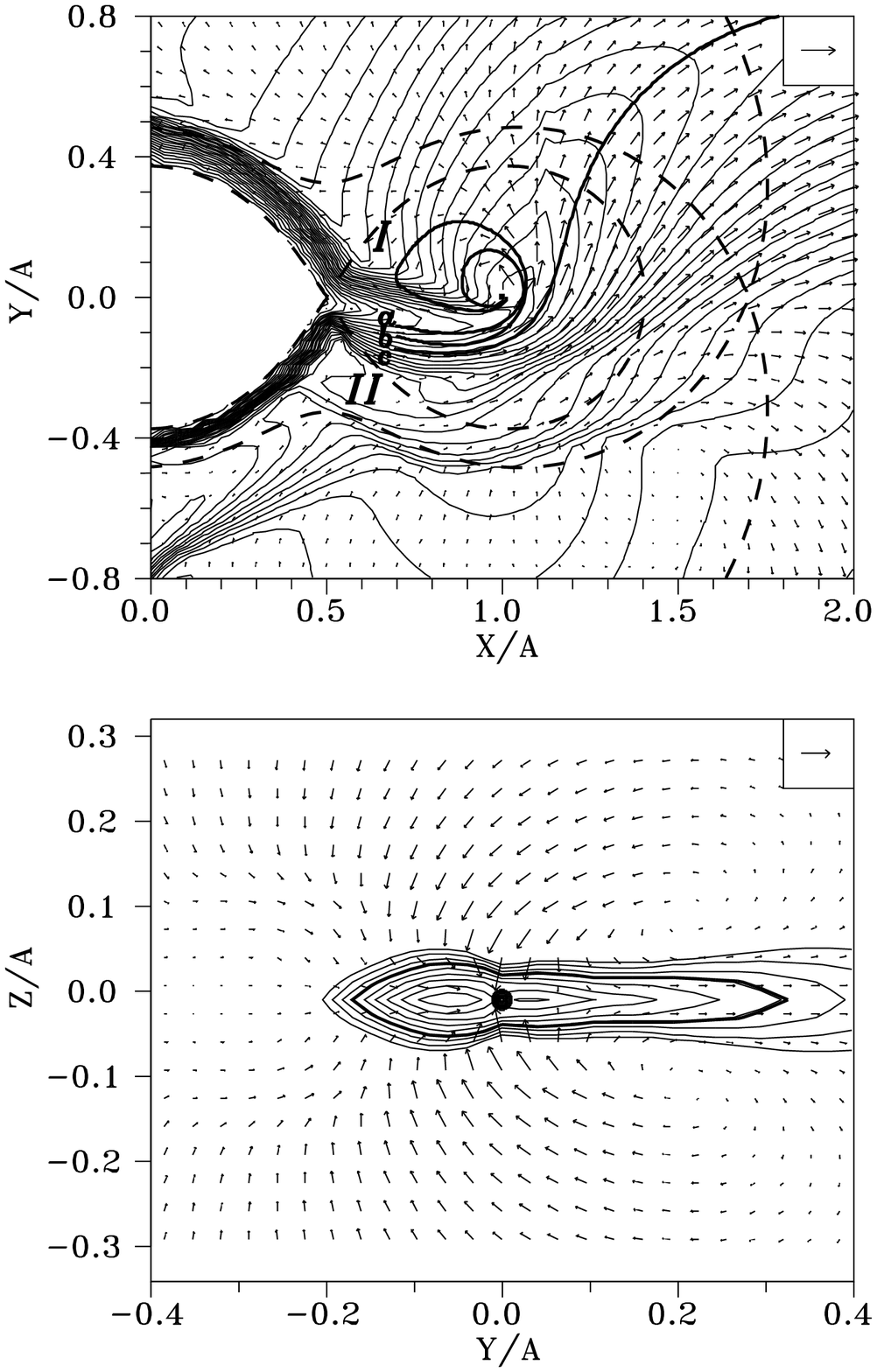}
\caption{Same as Figure~1 for calculation 5 from Table~I.}
\end{figure*}

\renewcommand{\thefigure}{6}
\begin{figure*}[p]
\plotone{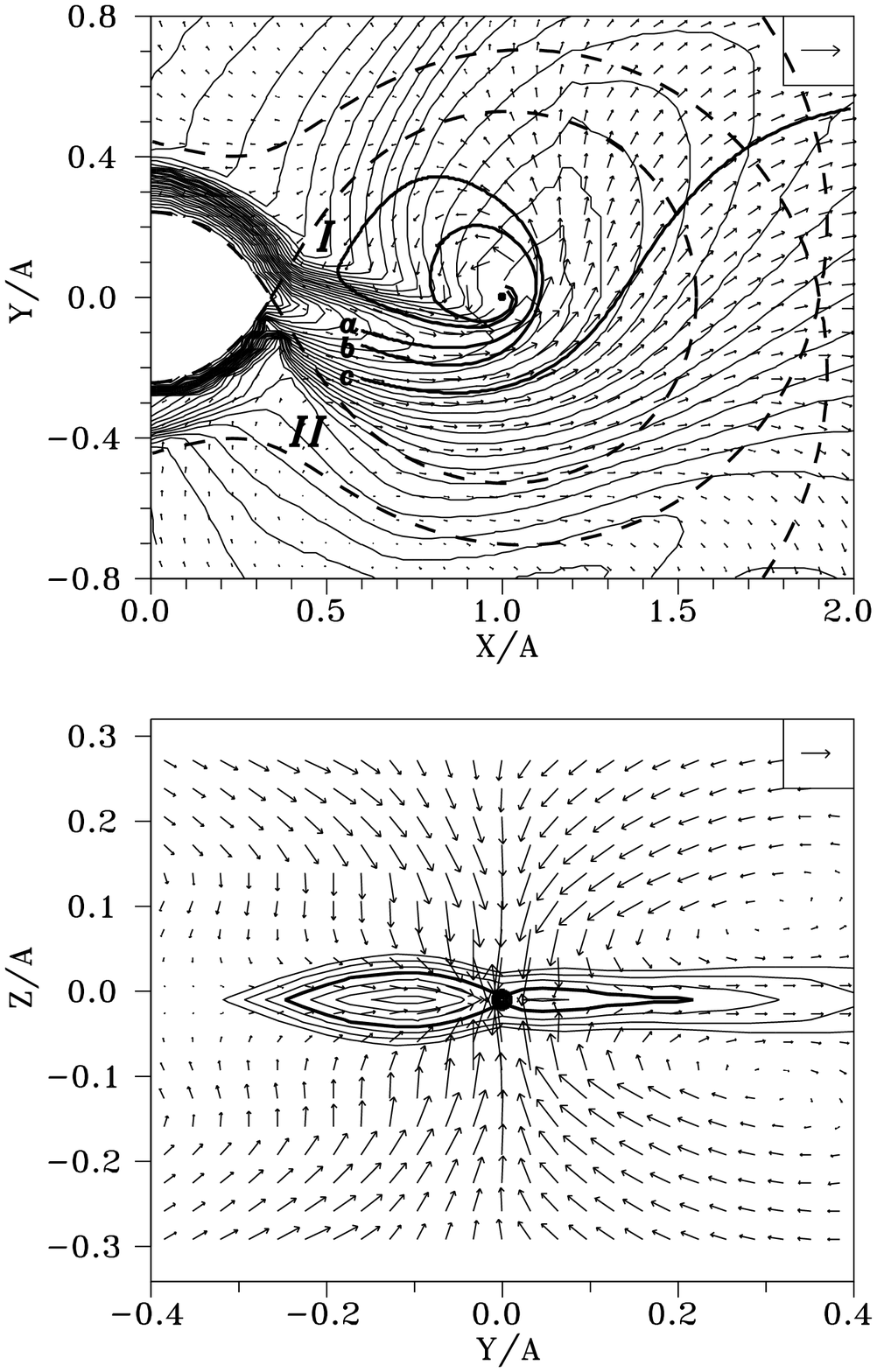}
\caption{Same as Figure~1 for calculation 6 from Table~I.}
\end{figure*}

\renewcommand{\thefigure}{7}
\begin{figure*}[p]
\plotone{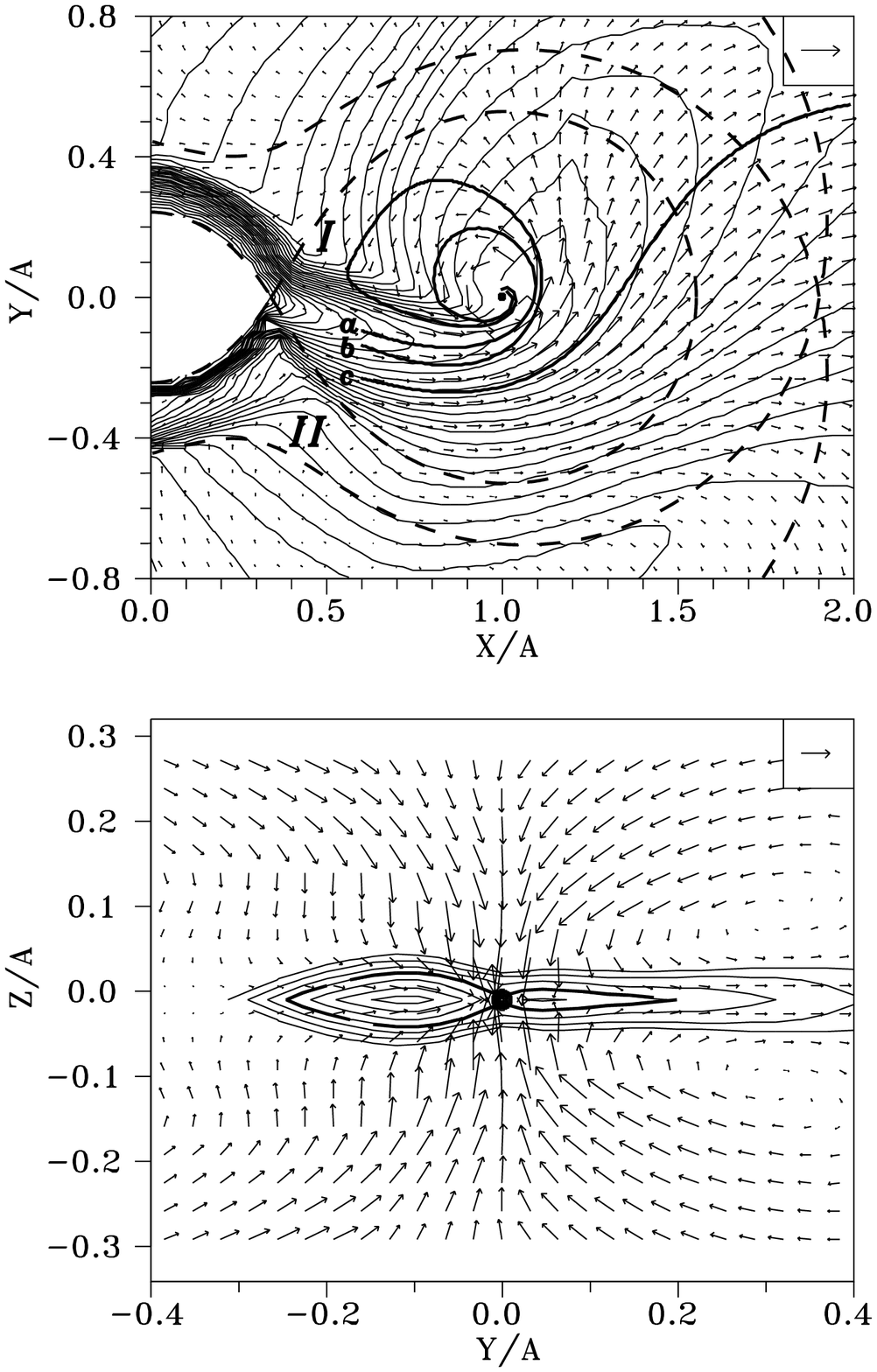}
\caption{Same as Figure~1 for calculation 7 from Table~I.}
\end{figure*}

\renewcommand{\thefigure}{8}
\begin{figure*}[p]
\plotone{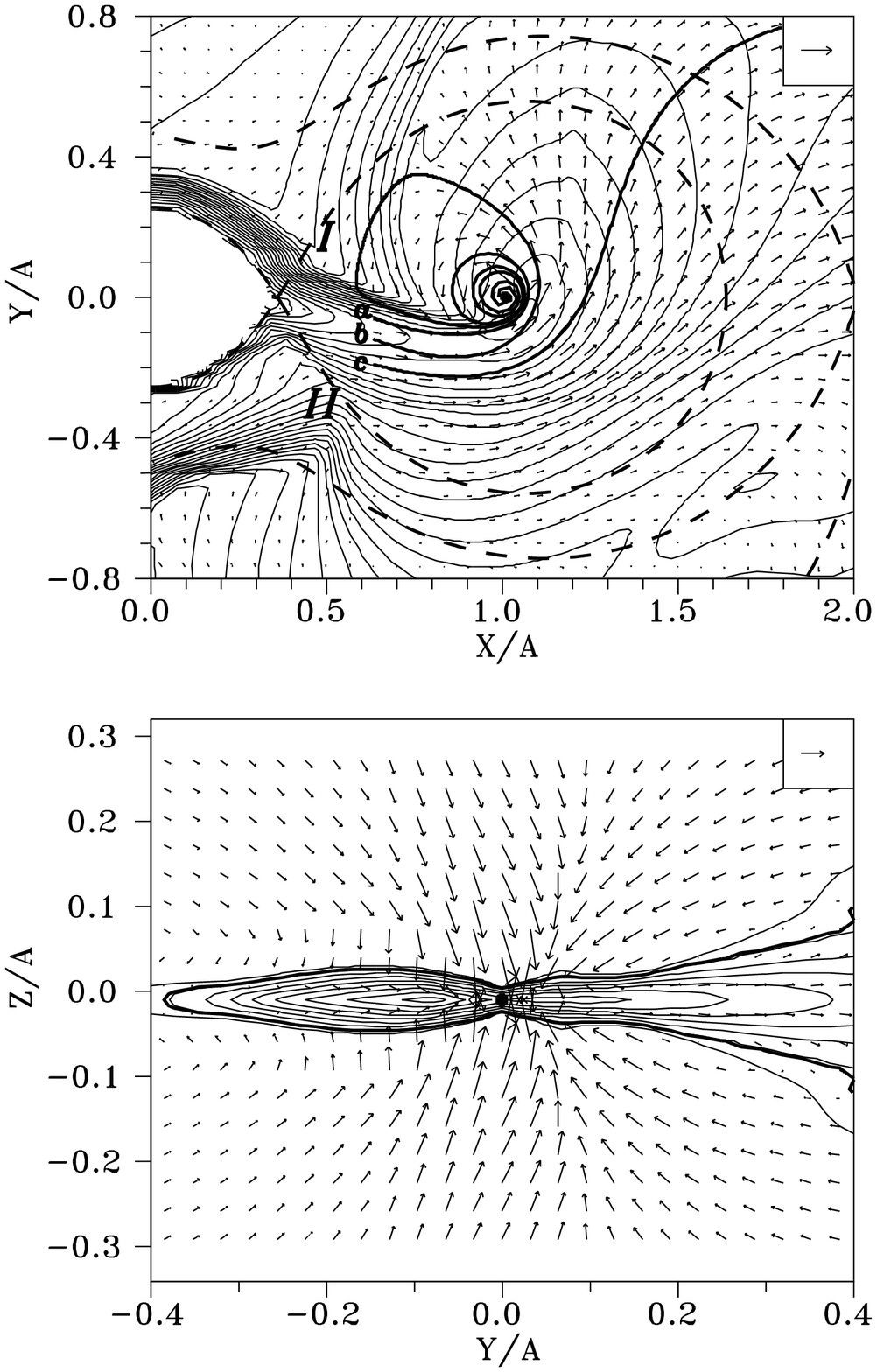}
\caption{Same as Figure~1 for calculation 8 from Table~I.}
\end{figure*}

    Taking into account the computer--intensive nature of the
problem, we included our previous results [22--25] in our
analysis. Additional computations were made in three cases
(computations 3, 4, 5 in Table~I), in order to fill out the
field of parameter values. We also used results presented by
other authors when their parameters differed from those in Table
I.

\bigskip
\begin{table}[ht]
\begin{center}
Table~I.\\
\vspace*{0.5cm}
\begin{tabular}{|c|c|c|c|c|}
\hline
Case N& $q$& $\epsilon$& ${\cal M}$& $\gamma$\\
\hline
1&1&0.15&0.083&1.2\\
\hline
2&1&0.15&0.083&1.01\\
\hline
3&1&0.016&0.083&1.2\\
\hline
4&1&0.016&0.083&1.01\\
\hline
5&1&0.016&1&1.01\\
\hline
6&5&0.016&0&1.01\\
\hline
7&5&0.016&1&1.01\\
\hline
8&5&0.04&1&1.01\\
\hline
\end{tabular}
\end{center}
\end{table}

    The computation results for each of the eight cases
presented in Table~I are given in Figures~1--8. The upper panels
(Figures~1a--8a) depict density contours and velocity vectors in
the equatorial plane of the system illustrating the morphology
of gas flows in the system. A vector showing the velocity scale
is given in the upper right corner of each plot. In all cases,
the length of this vector corresponds to the velocity $4 A \cdot
\Omega$. We also show in the figures three flowlines marked by
letters `$a$', `$b$', and `$c$', which illustrate the directions
of the flow of matter in the system. Density contours and
velocity vectors in the plane passing through the accretor and
perpendicular to the line connecting the centers of the two
components are shown in the lower panels (Figures~1b--8b).  The
velocity scale is specified by the vector in the upper right
corner of these panels, which corresponds to the velocity $2 A
\cdot \Omega$.  For convenience, the range of density variation
in Figures~1b--8b is chosen to be the same, and to be equal to
0.1 -- 0.0003${\rho}_0$. The thick line corresponds to the
density 0.001${\rho}_0$. The flow pattern in the figures is
stationary and corresponds to a steady-state flow regime.

    Analysis of the results presented in this paper enables
investigation of the influence of various parameters on the flow
structures in semidetached binary systems.

\subsection{The effect of $\gamma$}

    The dependence of the flow pattern on the adopted value of
the parameter $\gamma$ can be established by comparing the
results of calculations in cases 1 and 2, and in cases 3 and 4.
We varied only the parameter $\gamma$ in these pairs of cases,
with all other parameters having fixed values.

    Analysis of the results presented in the figures shows that,
in the quasi-stationary case (0.001 ${\rho}_0$), the flow
pattern is characterized by the presence of an accretion disk
and circumbinary envelope (Figures~2 and 4). The details of the flow
structure in the cases under consideration are largely
determined by the presence of the circumbinary envelope, and were
considered by us in [22--25]. Below, we give only a brief
description of the main features of the flow for this case,
referring the reader to [22--25] for a more detailed discussion.
Analysis of the gas flows allows us to divide the matter in the
stream flowing from $L_1$ into three parts: the first produces a
quasi-elliptical accretion disk (flowline `$a$') and continues
to participate in the accretion process, losing angular momentum
under the action of viscosity; the second (flowline `$b$')
flows around the accretor outside the disk; the third part of
the stream moves toward the Lagrange point $L_2$ (flowline
`$c$'), and a significant portion of this matter changes the
direction of its motion under the action of the Coriolis force
and remains in the system. The matter remaining in the system
and not involved directly in the accretion process (flowlines
`$b$' and `$c$') forms the circumbinary envelope of the system. A
significant portion of the envelope gas interacts with
the matter flowing from the surface of the donor star, and
substantially changes the mass exchange regime in the system
[23, 25]. Another part of the circumbinary envelope (flowline `$b$')
passes around the accretor and undergoes a shock interaction
with the edge of the jet facing the direction of the orbital
motion. This interaction leads to the absence of a shock
interaction of the stream flowing from $L_1$ with the gas of the
forming disk. The stream, deflected by the envelope gas,
comes at the disk along a tangent, and does not produce a shock
disturbance at the disk edge ("hot spot"). At the same time, the
interaction of the stream with the circumbinary envelope results in
the formation of an extended shock wave (denoted $I$ in the
figures) along the edge of the stream, which can be considered
equivalent to a "hot spot" on the disk in terms of its
observational manifestation [22, 24]. The interaction of the
circumbinary envelope with the material in the stream also leads to
the formation of shock wave $II$.

    The flow pattern undergoes qualitative changes in
calculations with the higher index $\gamma=1.2$ (Figures~1 and 3).
This is revealed primarily in the absence of the accretion disk.
Analysis of the flowlines presented in Figures~1a and 3a shows
that some portion of the stream material (flowline `$a$') flows
directly onto the accretor, whereas another portion (flowline
`$b$') passes around the accretor and undergoes a shock
interaction with the stream flowing from $L_1$, without forming
a disk. This fact is supported by the appearance of the density
and velocity-vector fields in the $YZ$ plane in Figures~1b and
3b.  Comparison of these with Figures~2b and 4b provides
conclusive evidence for the absence of an accretion disk for the
higher $\gamma$ value.  Note that this result is in agreement
with [14, 16], which demonstrated that a disk forms only when
$\gamma$ is close to unity ($\gamma < 1.1$). Additional
distinctions in the flow pattern for the different $\gamma$
reveal themselves in changes of the directions of flows in the
system. In particular, analysis of the gas motions in the $Z$
direction shows that, for the higher $\gamma$, the gas moves
predominantly away from the equatorial plane (Figures~1b and 3b),
whereas for $\gamma \sim 1$, it moves from higher heights toward
the accretion disk (Figures~2b and 4b).

   Along with the qualitative differences in the flow pattern,
the variations in the value of $\gamma$ give rise to significant
quantitative changes. In particular, for the low value of
$\gamma$, the influence of the circumbinary envelope on the surface of
the donor star substantially increases. This leads to an
increase in the rate of mass loss by the donor star and, as a
consequence, to an increase in the matter accretion rate by a
factor of 1.5--2.

\subsection{The effect of ${\cal M}$}

   Pairs of calculations that can be used to assess the role of
${\cal M}$ are denoted in Table~I by the numbers 4, 5 and 6, 7.
To study the role of the parameter ${\cal M}$, we compared
calculations in which the gas velocity at the near-surface layer
of the donor star was taken to be small (the Mach number ${\cal
M}=0.08$ in calculation 4 and the Mach number ${\cal M}=0$ in
calculation 6) with calculations in which the gas velocity
component normal to the surface was equal to the local sound
speed (${\cal M}=1$ for models 5 and 7 in Table~I).

    The results indicate that a number of minor quantitative
changes arise in the flow pattern. In particular, as the Mach
number increases, we observe a growth in the rate of mass loss
by the donor star due to the "stripping" of its stellar
atmosphere by the envelope gas. This effect is quite
natural, since an increase in the near-surface gas velocity is
accompanied by an increase in the overall extent of the
atmosphere, which leads to more intense stripping of atmospheric
material by the gas of the circumbinary envelope. Note, however, that
the growth of the mass-loss rate of the donor star does not
influence the accretion rate, which remains virtually constant as the Mach
number varies.

    At the same time, the general flow pattern retains all the
previously determined qualitative characteristics (see Figures~4
and 5 and also Figures~6 and 7). Moreover, the basic parameters of
the gas flows, dimensions and position of the accretion disk,
and position and intensity of shock waves $I$ and $II$ remain
virtually unchanged. This, in turn, suggests that ${\cal M}$ only
weakly influences the flow structure.

\subsection{The effect of $\epsilon$}

   The parameter $\epsilon=c_1/(\Omega A)$ was first introduced
by Lubow and Shu [26], and was used as a small parameter. Let us
estimate the characteristic range of this parameter for various
types of binary systems. Combining the expressions $\Omega
A=\sqrt{GM/A}$ and $c_1=\sqrt{\gamma R T_1}$, we can represent
$\epsilon$ as

$$
\epsilon=0.003 T_1^{1/2} P_{год}^{1/3} M_{M_\odot}^{-1/3}
{\gamma}^{1/2}\,.
$$
Here, the orbital period $P$ is expressed in years and the total
mass $M$ of the system in solar masses $M_\odot$. Let us assume
that the gas temperature at the donor star is $10^4$K. The
range of $\epsilon$ values characteristic of this case is
presented in Table~II for various $P$ and $M$. Analysis of the
data in Table~II shows that the range of $\epsilon$ is
rather small ($\sim 0.001 \div 0.3$). However, $\epsilon$
cannot always be considered a small parameter.

\bigskip
\begin{table}[ht]
\begin{center}
Table~II.\\
\vspace*{0.5cm}
\begin{tabular}{|c|c|c|c|c|c|}
\hline
\multicolumn{2}{|c|}{$M=M_\odot$}&\multicolumn{2}{c|}{$M=10M_\odot$}&\multicolumn{2}{c|}{$M=100M_\odot$}\\
\hline
$P$&$\epsilon$&$P$&$\epsilon$&$P$&$\epsilon$\\
\hline
year&0.3&year&0.13&year&0.065\\
\hline
month&0.13&month&0.05&month&0.028\\
\hline
day&0.04&day&0.02&day&0.009\\
\hline
hour&0.014&hour&0.006&hour&0.003\\
\hline
minute&0.0037&minute&0.0017&minute&0.0008\\
\hline
\end{tabular}
\end{center}
\end{table}

\renewcommand{\thefigure}{9}
\epsscale{1.0}
\begin{figure}[t]
\plotone{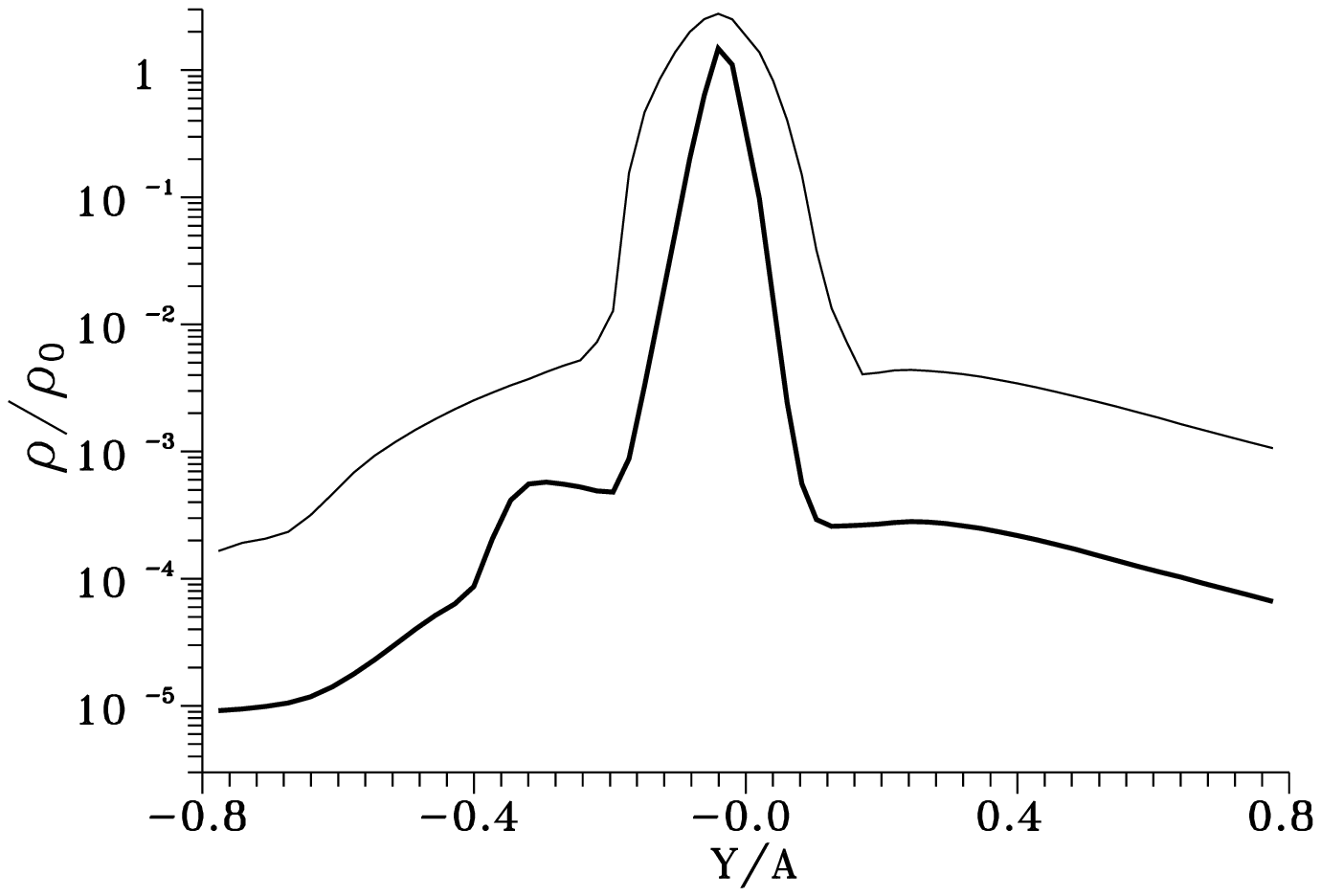}
\caption{One-dimensional density profiles for calculations 2 and
4 specified in Table~I. The density values are taken along the
straight line in the equatorial plane and passing through the
point $0.6A$ perpendicular to the line connecting the centers of
the components of the system.}
\end{figure}

   To determine the effect of the parameter $\epsilon$, we
compare calculations 2 and 4, and 7 and 8. In the first pair of
calculations, we varied $\epsilon$ from 0.15 to 0.016, and in
the second pair, from 0.016 to 0.04. Analysis of the results
(Figures~2, 4, 7, and 8) shows that the qualitative flow pattern
remains unchanged for various $\epsilon$: an accretion disk
forms in the system; there is no "hot spot" in the region of
interaction of the stream with the disk; the directions of the
flows coincide; and the forming circumbinary envelope interacts with
the stream, producing the shock waves $I$ and $II$. In addition,
a number of quantitative changes are observed as $\epsilon$ is
increased. In particular, the cross section of the stream
increases in the calculations with the larger $\epsilon$. This
is supported by Figure~9, which presents one-dimensional density
profiles for calculations 2 and 4. The density values are taken
along the straight line in the equatorial plane passing through
the point $0.6A$ perpendicular to the line connecting the
centers of the system's components.  This conclusion is
consistent with the results of Lubow and Shu [26], who showed
that, for small $\epsilon$, the width of the stream is
$\epsilon$. In addition to the increase in the stream size for
the larger value of $\epsilon$, we note an increase of the disk
thickness and the accretion rate.

\subsection{The effect of $q$}

   In this study, we examined systems in which the parameter
$q=M_2/M_1$ takes on only two values: $q=1$ in calculations 1--5
and $q=5$ in calculations 6--8.  Assessment of the role of this
parameter can be made using calculations 5 and 7, in which all
parameters except $q$ were the same. Comparison of the results
presented in Figures~5 and 7 shows that the qualitative flow
pattern remains unchanged, although some quantitative variations
are present. This indicates that the calculation results depend
only slightly on the adopted ratio of the masses of the
components when $q \geq 1$.

    The dependence of the flow pattern on $q$ when $q < 1$
requires additional study. According to [27, 28], when $q < 1$,
the flow pattern undergoes significant changes, which is
reflected, in particular, in clear observational manifestations.
In addition, considerable tidal disturbances are possible when
the mass of the donor star is large. A detailed study of the
influence of the parameter $q$ will be presented in a future
paper.

\section{Conclusion}

   Analysis of the 3D gas--dynamical calculations performed here
enables us to sort the parameters of the problem according to
their influence on the flow pattern in a semidetached binary
system. The variation in the adiabatic index $\gamma$ has the
largest effect on the solution. The influence of ${\cal M}$ and
$\epsilon$ on the flow pattern is small, and leads only to unimportant
quantitative variations of the solutions without changing the overall
qualitative flow patter. The ratio $q$ of the masses of the two
components also does not exert a significant influence for the
region $q \geq 1$.

    Since qualitative changes of the flow pattern depend only on
the adopted value of the parameter $\gamma$, let us consider the
mechanism of its action upon the flow and summarize the main
features of the gas-flow structure that should be observed in
semidetached binaries. We considered the flow of a nonviscous,
non-heat-conducting gas in a binary system without magnetic and
radiation fields. In the model, the assumption that the gas is
ideal formally implies neglect of radiative losses, although
small values of the adiabatic index $\gamma$ imitate energy
losses in the system (see, for example, [3]).  In particular,
when $\gamma \sim 1$, a solution close to the isothermic solution
is realized [28]. In the above calculations, we used the two
$\gamma$ values 1.2 and 1.01, and, thus, obtained solutions
corresponding to models with different energy-loss rates. The
flow patterns calculated in these two cases display fundamental
differences. For example, in the calculations performed with
$\gamma=1.2$, the accretion disk does not form at all. The
increase in energy losses in the system in the calculations for
the close to isothermal case ($\gamma \sim 1$) leads to the
formation of an accretion disk and a corresponding change in the
accretion mechanism.

    In typical binaries, the gas between the components is
ionized, and the corresponding energy-loss mechanisms work
efficiently, so the real flow pattern should be described by a
close to isothermal model [30, 31]. As we noted above, in this
case, the flow pattern is characterized by the following
features: the stream of gas flowing from $L_1$ forms both an
accretion disk and a circumbinary envelope in the system. Further, the
envelope gas exerts a considerable influence on the
structure of gas flows in the system [23--25].  In particular,
the envelope gas interacts with the stream of matter flowing
from the vicinity of $L_1$ and deviates it, giving rise to a
shockless (tangential) interaction of the stream with the outer
edge of the forming accretion disk and, therefore, to the
absence of a "hot spot" in the disk. At the same time, the
interaction of the envelope gas with the stream is
responsible for the formation of an extended shock wave at the
edge of the stream, whose observational manifestations can be
considered equivalent to those of a hot spot in the disk [24].

\acknowledgments

This work was supported by the Russian Foundation for Basic Research
(project code 96-02-16140), the State Scientific and Technological Program
"Astronomy" ("Computational Astrophysics" Section), and the "Cosmion"
Scientific Educational Center for Cosmoparticle Physics.

\end{document}